# Photonic spiking reinforcement learning for intelligent routing

Shuiying Xiang[1*], Yonghang Chen[1], Ling Zheng[1,2*], Zhicong Tu[1], Xintao Zeng[1], Mengting Yu[1],

Shuai Wang[1], Yahui Zhang[1], Xingxing Guo[1], Weitao Pan[1*], Yue Hao[1]

[1] State Key Laboratory of Integrated Service Networks, Xidian University, Xi'an 710071, China;
[2] School of Communication and Information Engineering, Xi'an University of Posts and Telecommunications, Xi'an 710121, China
*Correspondence: syxiang@xidian.edu.cn, lingzheng@xupt.edu.cn, wtpan@mail.xidian.edu.cn

**Abstract**: Intelligent routing plays a key role in modern communication infrastructure, including data centers, computing networks, and future 6G networks. Although reinforcement learning (RL) has shown great potential for intelligent routing, its practical deployment remains constrained by high energy consumption and decision latency. Here, we propose a photonic spiking RL architecture that implements a proximal policy optimization (PPO)–based intelligent routing algorithm. The performance of the proposed approach is systematically evaluated on a software-defined network (SDN) with a fat-tree topology. The results demonstrate that, under various baseline traffic rate conditions, the PPO-based routing strategy significantly outperforms the conventional Dijkstra algorithm in key performance metrics, including throughput, packet loss rate, average latency, and load balance. Furthermore, a hardware-software collaborative framework of the spiking Actor network is realized for three typical baseline traffic rates, utilizing a photonic synapse chip based on a Mach-Zehnder interferometer (MZI) array and a photonic spiking neuron chip based on distributed feedback lasers with a saturable absorber (DFB-SAs). Experimental validation on 640 state–action pairs shows that the inference accuracy of the hardware-software collaborative framework is consistent with that of the pure algorithmic implementation. The impacts of different hidden-layer scales in the spiking Actor network and varying network size of fat-tree topology are further analyzed. The integration of photonic spiking RL with SDN-based routing establishes a novel paradigm for intelligent routing optimization, featuring ultra-low latency and high energy efficiency. This approach exhibits broad application prospects in real-time network optimization scenarios, including large-scale data centers, computing networks, satellite Internet systems, and future 6G networks.

## 1. Introduction

Emerging information technologies, including cloud computing, big data, artificial intelligence, and the Internet of Things, are reshaping social production and lifestyle at an unprecedented speed. This transformation is accompanied by explosive data growth and increasingly stringent real-time processing demands, imposing exceptionally high requirements on the performance, efficiency, and intelligence of underlying network infrastructures. Routing technology, which is responsible for resource allocation optimization, quality-of-service assurance, and overall efficiency enhancement, has become a critical research frontier in information and communication technologies [1].

Traditional routing algorithms, exemplified by the shortest-path–based Dijkstra algorithm and its implementations in protocols such as open shortest path first (OSPF), have long served as the foundation of Internet routing [2]. However, these approaches rely on static or quasi-static deterministic decision-making mechanisms driven by local or fixed metrics (e.g., hop count and link bandwidth). When confronted with highly dynamic, bursty, and heterogeneous traffic patterns in modern networks, particularly in data center and computing power networks, this paradigm exhibits inherent limitations. Firstly, they lack real-time global state awareness and historical learning capabilities. Secondly, their fixed-rule strategies are too rigid to adaptively learn and respond to unknown or complex nonlinear traffic patterns. Finally, their single-objective optimization (e.g., minimizing hop count) often leads to local congestion and inefficient resource utilization.

To overcome the limitations of traditional routing approaches, deep reinforcement learning (DRL) has emerged as a promising paradigm for intelligent routing optimization [3-7]. DRL agents can learn complex dynamic characteristics from high-dimensional state spaces and converge toward globally optimized routing strategies by continuously interacting with network environments [8-12]. Among various DRL algorithms, the proximal policy optimization (PPO) algorithm has demonstrated strong potential in complex routing optimization tasks due to its training stability, high sample efficiency, and support for both continuous and discrete action spaces [13]. Nevertheless, DRL-based routing with conventional artificial neural networks (ANNs) still suffers from high computational energy consumption and decision latency, conflicting with the ultra-low latency and resource-constrained requirements of practical network control planes—especially in edge computing scenarios.

Spiking neural networks (SNNs), inspired by biological neural information processing, provide an attractive alternative for improving energy efficiency and temporal dynamics compared to ANN-based models [14, 15]. Meanwhile, photonic computing exploits the intrinsic physical advantages of light, including high speed, massive parallelism, ultra-wide bandwidth, and high energy efficiency, thereby offering a powerful hardware platform for neuromorphic computing [16, 17]. By integrating SNN computational principles with photonic hardware, photonic spiking neural networks (PSNNs) have emerged as a key direction for next-generation high-performance and energy-efficient intelligent computing architectures [18, 19]. PSNNs based on distributed feedback lasers with saturable absorbers (DFB-SAs) chips have already demonstrated preliminary successes in representative tasks such as image classification [20, 21]. In recent years, photonic chips have been extended to reinforcement learning–related applications [22, 23]. Jin et al. developed a programmable hybrid photonic chip capable of high-precision optical dot-product operations with dimensionality up to 15, enabling the execution of complex RL algorithms on photonic platforms. However, research on integrating this emerging technology with the critical task of intelligent routing remains largely unexplored.

In this work, we propose a photonic spiking reinforcement learning–based intelligent routing architecture for software-defined networking (SDN) [24, 25], which tightly integrates the powerful learning and optimization capability of the PPO algorithm with the ultra-low latency and high energy efficiency of PSNN hardware. This study aims to bridge the research gap that exists at the intersection of photonic neuromorphic computing and network routing technologies. The main contributions of this work can be summarized as follows:

**(1) Architecture and algorithm innovation.** An intelligent routing model based on the PPO framework is developed, in which the policy network responsible for routing decisions is implemented using an SNN. This design aligns the computational paradigm of the routing agent with the characteristics of photonic neuromorphic hardware, enabling efficient hardware deployment. The proposed model is integrated into the SDN control plane and evaluated on a representative fat-tree data center topology.

**(2) Performance improvement.** Extensive comparative experiments are conducted in a simulated fat-tree network to evaluate the proposed photonic spiking PPO routing algorithm against the conventional Dijkstra algorithm. Comprehensive assessments across throughput, latency, packet loss rate, and load balance consistently validate the superior overall performance of our approach.

**(3) Hardware-software collaborative computing.** A hardware-software collaborative computing framework is developed for mapping the spiking Actor network onto PSNN hardware. Experimental validation on 640 representative state–action pairs confirms that the decision accuracy and algorithmic performance are fully preserved during the mapping from software simulation to hardware execution.

The remainder of this paper is organized as follows. Section II introduces the overall architecture and operating principles of the proposed photonic spiking reinforcement learning–based intelligent routing framework, describes the experimental platform used for hardware-software collaborative computing, and systematically details the design, implementation, and training methodology of the photonic spiking PPO routing algorithm. Section III presents and analyzes the simulation results obtained under a fat-tree network topology. Section IV reports the hardware-software collaborative computing results. Section V investigates the impacts of different SNN hidden-layer scales and varying fat-tree network sizes. Finally, Section VI concludes the main findings. This work establishes a novel paradigm for intelligent routing optimization characterized by ultra-low latency, high energy efficiency, and enhanced intelligence, offering significant value for future large-scale data centers, computing networks, satellite Internet systems, and future 6G networks.

## 2. Overall architecture of photonic spiking PPO for intelligent routing

### 2.1 Overall architecture

The proposed photonic spiking reinforcement learning–based SDN intelligent routing system adopts a hierarchical architecture consisting of a data plane, a control plane, and an application plane (intelligent decision-making layer), as illustrated in Fig. 1. The data plane adopts a fat-tree topology with a parameter of k = 4 and is composed of multiple programmable OpenFlow-enabled switches. It is responsible for real-time packet forwarding and continuous acquisition of network state information. Key metrics, including link bandwidth, latency, and load conditions, are collected and reported to the control plane via the OpenFlow protocol. Meanwhile, the data plane executes routing decisions by installing flow table rules issued by the control plane. In this study, the fat-tree network comprises 16 hosts and 20 switches, with a uniform link capacity of 1000 Mbps. During training, the baseline traffic rate is set to 50 Mbps, and the traffic pairs are fixed at 64. For performance evaluation, the baseline traffic rate varies from 40 Mbps to 200 Mbps to comprehensively assess routing performance under different network load conditions.

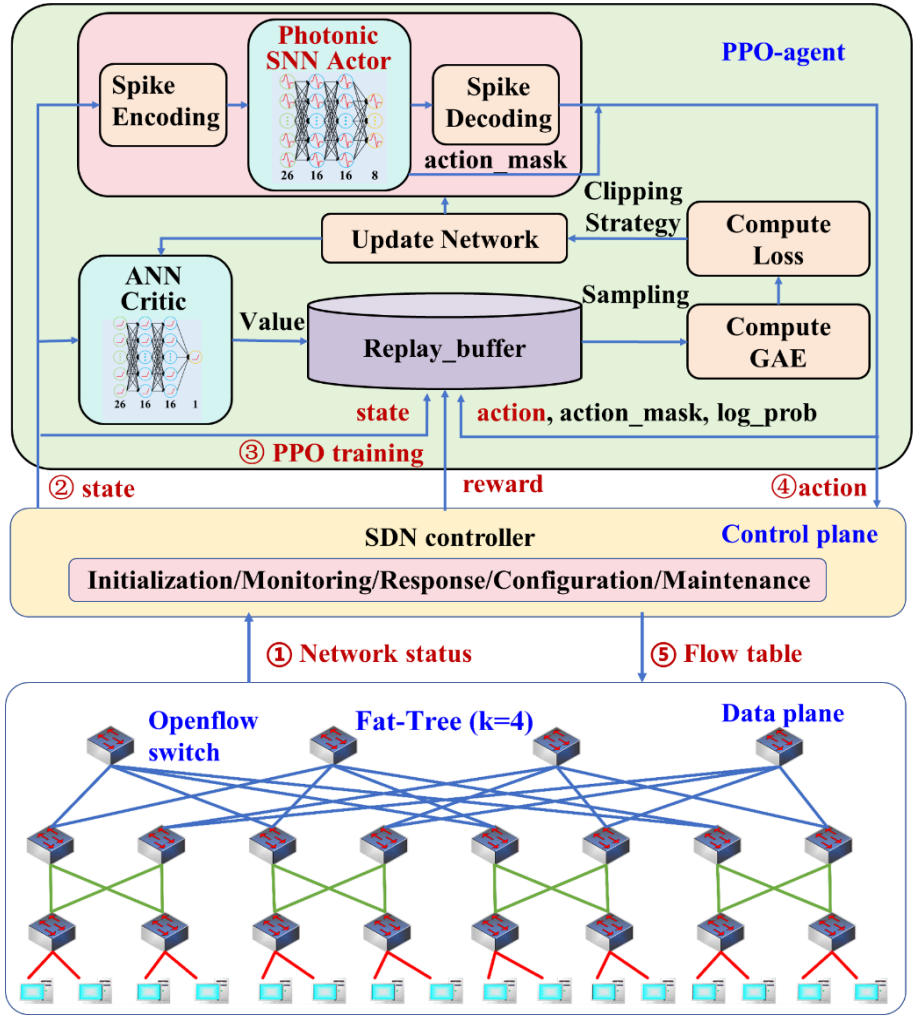

Fig. 1 Overall architecture of photonic spiking PPO for intelligent routing in SDN.

The control plane is built upon an SDN controller and provides centralized network management and routing enforcement. It consists of five functional modules: **Initialization**, which registers switches and establishes control channels; **Monitoring**, which periodically collects network state information (e.g., link utilization and packet statistics); **Response**, which processes Packet-In messages and triggers routing decisions; **Configuration**, which installs flow table rules generated by the intelligent decision-making layer; and **Maintenance**, which manages flow table timeouts and network state updates.

The application plane serves as the core innovation of the system and is implemented using a photonic spiking reinforcement learning engine based on the PPO Actor–Critic framework. It processes network state information and generates optimized routing strategies. The photonic spiking Actor network is realized using an SNN and consists of three modules: Spike Encoding, SNN Layer, and Spike Decoding. Network states are encoded into spike trains, processed by the SNN to generate routing actions, and decoded with an action mask to filter invalid decisions. The Critic network is implemented using an ANN to evaluate the state value and assist policy optimization. The PPO training module incorporates a replay buffer, generalized advantage estimation (GAE), loss computation, clipping strategy, and network update mechanisms, enabling stable and efficient policy optimization.

The spiking Actor network employs leaky integrate-and-fire (LIF) neurons as its fundamental computational units, inspired by biological neural dynamics. At each discrete time step $t$, the membrane potential of the neuron $u_t$ is updated according to Eqs. (1) and (2):

$$u_{t+1} = \tau \cdot u_t + x_t - V_{th} \cdot S(u_t) \tag{1}$$

$$S(u_t) = \begin{cases} 1, & \text{if } u_t \geq V_{th}; \\ 0, & \text{if } u_t < V_{th} \end{cases} \tag{2}$$

Where the input current (or input signal) is $x_t$, the spike firing threshold is $V_{th}$, the leakage coefficient $\tau$ ($0<\tau<1$) controls the temporal memory of the membrane potential, and a spike is emitted when the potential exceeds a predefined threshold, followed by a membrane reset. To address the non-differentiability of the spike firing function $S(u_t)$, a zero-inflated firing (ZIF) surrogate gradient method is adopted to approximate the spike activation function, enabling gradient-based backpropagation as defined in Eq. (3). Furthermore, gradient clipping is introduced during training to stabilize membrane potential updates and ensure effective optimization of the SNN, as described in Eq. (4).

$$S(u_t) \approx \sigma(u_t) = \begin{cases} 1, & \text{if } u_t > 0; \\ 0, & \text{if } u_t \leq 0 \end{cases} \tag{3}$$

$$\frac{\partial \sigma(u_t)}{\partial u_t} \approx clip(-1,+1) \tag{4}$$

**2.2 Intelligent routing optimization process**

Based on the proposed architecture, the intelligent routing optimization process is organized into five sequential stages as follows:

**Stage 1: Network state collection.** OpenFlow-enabled switches in the data plane continuously monitor network conditions, including topology, link utilization, latency, traffic distribution, and packet loss, etc. The collected information (①Network Status) is reported to the SDN controller in the control plane via the southbound interface.

**Stage 2: State perception and representation.** The SDN controller normalizes and aggregates the collected network information to form a reinforcement learning–compatible state representation(②state), which is then forwarded to both the Actor and Critic networks of the PPO agent.

**Stage 3: PPO training and policy optimization.** This stage constitutes the core optimization process of the photonic spiking reinforcement learning framework(③PPO training). First, the Spike Encoding module converts the input state into spike trains, which are processed by the spiking Actor network consisting of multiple fully connected layers (26×16×16×8). After spike decoding, the Actor outputs routing actions and their corresponding log probabilities, along with an action mask to eliminate invalid actions. Meanwhile, the Critic network evaluates the input state and outputs its value for advantage estimation. The tuples {state, action, action mask, action probability, reward} are stored in a replay buffer. During training, experiences are sampled to compute the GAE, and the Actor and Critic networks are iteratively updated using the PPO loss function with a clipping strategy.

**Stage 4: Action decision-making and flow table generation.** After convergence, the trained Actor network generates the optimal routing action for the current network state (④action), which is transmitted to the SDN controller in the control plane.

**Stage 5: Flow table issuance and routing execution.** The SDN controller translates the routing decision into flow table entries and installs them on the OpenFlow switches in the data plane(⑤Flow table). The switches execute packet forwarding based on the updated flow tables, thereby completing the closed-loop intelligent routing process.

**2.3 Hardware-software collaborative training-inference framework**

Photonic hardware computing is inherently susceptible to fabrication imperfections and system noise, resulting in potential accuracy degradation. To mitigate both SNN training challenges and optical inference errors, we adopt an end-to-end hardware-software collaborative training–inference framework for PSNNs, following a similar methodology to Ref. [26].

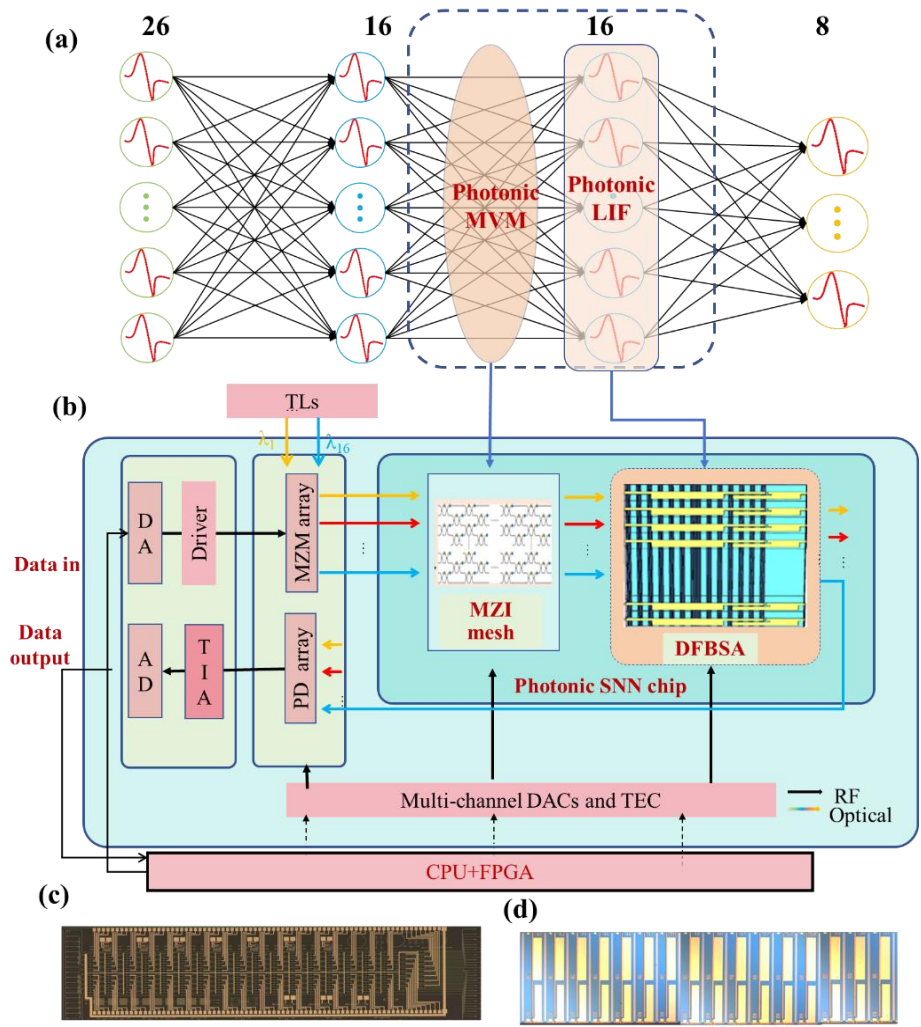

Fig.2 Schematic diagram of optoelectronic collaborative computing setup. (a) SNN architecture; (b) photonic hardware implementation of SNN; (c) MZI mesh chip [26]; (d) DFB-SA array chip [26].

**Step 1: Software pre-training.** The Actor network employs an SNN trained from scratch using the single-time-step (T=1) surrogate gradient method, whereas the Critic is implemented as an ANN. In Actor–Critic–based DRL algorithms, the Critic network is not involved in hardware inference and therefore remains implemented using an ANN, enabling accurate estimation of discounted cumulative rewards and stable policy learning.

**Step 2: Photonic hardware in-situ training of MZI chip.** The 16×16 weight matrix of the hidden-layer linear transformation in the pre-trained spiking Actor network is exported and mapped on the MZI mesh chip through online training. To map the digital weights onto the photonic hardware, the stochastic parallel gradient descent (SPGD) algorithm is employed to optimize the phase-shifter heater voltages of the MZI mesh [27]. The SPGD-based training framework iteratively performs random voltage perturbation, gradient estimation, and voltage update. In simulation, the converged hidden-layer weight matrix obtained from Step 1 is used as the target matrix. Through iterative SPGD optimization, a voltage configuration table for the MZI phase shifters is obtained, yielding the corresponding hardware-equivalent weight matrix and completing the mapping from digital-domain weights to optical-domain implementation.

**Step 3: Hardware-aware software fine-tuning.** After MZI training, the obtained weight matrix is reloaded into the algorithm for retraining to compensate for fabrication- and noise-induced accuracy loss. The intermediate linear-layer weights are fixed with gradients disabled. After retraining, the relevant layer inputs and outputs are exported for hardware–software collaborative inference. Hardware validation is conducted under three load conditions (40, 80, and 120 Mbps) using 640 × 16 state–action sample pairs.

**Step 4: Hardware-aware collaborative inference.** The optoelectronic collaborative computing framework is illustrated in Fig. 2 (b). Owing to current PSNN chip scale constraints, only the intermediate hidden layers are deployed on the photonic platform, including photonic matrix–vector multiplication (MVM) and photonic LIF neuron modules. Specifically, the nonlinear spiking activation output of the previous 26×16 network, computed on the CPU–FPGA platform, is used as the input to the photonic computing network and is transmitted to the digital-to-analog (DA) converters and driver circuits. These electronic signals are modulated onto optical carriers generated by a tunable laser array using a Mach–Zehnder modulator (MZM) array. The modulated optical signals are then processed by an MZI mesh to perform linear MVM operations, followed by nonlinear spiking activation implemented by a DFB-SA array. The resulting optical spike signals are converted back to the electronic domain using photodetectors (PDs), amplified by transimpedance amplifiers (TIA), and digitized via analog-to-digital (AD) converters, completing one layer of optoelectronic hybrid computation before feeding into subsequent layers. During experiments, the MZI mesh and DFB-SA array are precisely driven and stabilized using programmable multi-channel voltage sources and high-precision temperature control, ensuring reliable and repeatable photonic spiking inference. The recorded outputs are then returned to the third linear layer of the algorithm for the next layer of computation. Finally, the obtained results are compared item by item with those of the fine-tuned software model to calculate the hardware-software collaborative computing accuracy.

**2.4 Photonic spiking PPO algorithm**

To implement the proposed photonic spiking PPO algorithm, a simulation environment consistent with practical SDN scenarios is constructed, including controllers, switches, and hosts. The customized environment defines the state as network characteristics, the action as path selection, and the reward as a joint function of throughput, packet loss rate, average delay, and load balance.

**(1) State space.** The state space is designed to capture key information required for routing decisions and consists of host encoding and path features. The host encoding includes normalized indices of the source and destination hosts $[src_{norm}, dst_{norm}]$, enabling the agent to distinguish different traffic pairs. The path feature component includes up to 8 candidate paths, each characterized by 3 features: average utilization $U_{avg}(p)$, maximum utilization $U_{max}(p)$, and normalized hop count $H_{norm}(p)$. Together, these form a 26-dimensional state vector:

$$s = [src_{norm}, dst_{norm}, U_{avg}(p_1), U_{max}(p_1), H_{norm}(p_1), ..., U_{avg}(p_8), U_{max}(p_8), H_{norm}(p_8)] \quad (5)$$

where the detailed definitions are given by:

$$\begin{cases} src_{norm} = \frac{index(src)}{N_{hosts}}, dst_{norm} = \frac{index(dst)}{N_{hosts}}, U_{avg}(p) = \frac{1}{|E_p|} \sum_{(u,v) \in E_p} \frac{L(u,v)}{C} \\ U_{max(p)} = max_{(u,v) \in E_p} \frac{L(u,v)}{C}, H_{norm}(p) = \frac{|P|-1}{10} \end{cases} \quad (6)$$

Where $index(\cdot)$ denotes host indices; $N_{host}$ is the total number of hosts; $E_p$ represents the edge set of path $p$; $L(u, v)$ is the current load of link $(u, v)$; $C$ is the link capacity (1000 Mbps); and $|P|$ denotes the hop count of path $p$ (i.e., the number of links that path $p$ traverses).

**(2) Action space.** The action space is defined as a discrete path-selection vector $\vec{a} = [a_0, a_1, ..., a_m]$, $m = min$ (actual number of paths, 8), where the number of valid actions varies dynamically according to available paths. An action mask is employed to ensure that only feasible paths can be selected:

$$mask[i] = \begin{cases} 1 & \text{if } i < m \text{ (path exists)} \\ 0 & \text{if } i \geq m \text{ (path doesn't exist)} \end{cases} \quad (7)$$

The mask is applied to the logits output of the Actor network to obtain the effective action probability distribution $p_i$:

$$\begin{cases} p_i = \dfrac{\exp(\text{logits}'[i])}{\sum_{j=0}^{m-1} \exp(\text{logits}'[j])}, \\ \text{logits}'[i] = \begin{cases} \text{logits}[i] & \text{if mask}[i] = 1 \\ -\infty & \text{if mask}[i] = 0 \end{cases} \end{cases} \quad (8)$$

**(3) Reward.** The reward function jointly optimizes multiple optimization objectives such as transmission efficiency, reliability, and load balance, and is defined as:

$$R_{total} = R_{base} + R_{traffic} - P_{drop} - P_{load} \times W_{factor} - P_{hop} \times W_{hop} + R_{balance} \quad (9)$$

Where $R_{base} = 30.0$ provides a stable positive reward baseline; $R_{traffic}$ encourages high throughput; $P_{drop}$ penalizes packet loss with increasing severity under higher traffic loads; $P_{load}$ penalizes uneven link utilization, weighted by the network utilization factor $W_{fActor}$; $P_{hop}$ discourages long paths through hop-count penalties; and $R_{balance}$ promotes long-term load balancing based on the balancing degree $B_{long}$. The expanded reward formulation is given in

$$R = 30.0 + 0.3D - \left(1 - \dfrac{D}{T}\right)(1.5T + 20.0) - \left(50U_{avg}^2 + 60U_{max}^2\right) \times (0.8 + 1.2U_{net}) \\ - 2.0H \times (1.2 - \min(U_{net}, 0.8)) + 50.0B_{long} \quad (10)$$

Where $D$ is the successfully transmitted traffic (in Mbps), $T$ is the target transmission volume of the traffic (in Mbps), $U_{avg}$ denotes the average utilization of the links traversed by path $p$, $U_{max}$ denotes the maximum utilization of the links traversed by path $p$, $U_{net}$ is the overall link utilization of the entire network, $H$ is the hop count of the path, and $B_{long}$ represents the long-term load balancing degree. Finally, reward values are clipped to the range [−40.0, 100.0] to prevent reward explosion.

To evaluate the performance of the intelligent routing algorithm, four key metrics are adopted.

**(1) Throughput.** Throughput measures the amount of data successfully transmitted per unit time, reflecting the overall data-carrying capacity of the SDN network. It is influenced by link bandwidth, forwarding capability, and traffic scheduling, and is defined as

$$\text{Throughput} = \dfrac{\sum_{i=1}^{N} ap_i \times 10^6 \times \Delta t}{\Delta t \times 10^6} = \dfrac{\sum ap_i \times 10^6}{10^6} \text{Mbps} \quad (11)$$

Where $ap$ denotes the total successfully transmitted data (in bits) within the statistical period $\Delta t$ (in seconds).

**(2) Packet loss rate.** Packet loss rate quantifies the ratio of packets that fail to reach the destination to the total transmitted packets within a statistical period, indicating network reliability. It is typically caused by congestion, insufficient resources, or link failures, and is calculated as

$$\text{PacketLossRate} = \left(1 - \frac{dp}{tp}\right) \times 100\%, \quad tp = \frac{tm \times 10^6 \times \Delta t}{1500 \times 8} \tag{12}$$

Where $dp$ is the total number of successfully received packets, $tp$ is the total number of transmitted packets, and $tm$ denotes the network transmission rate (in Mbps).

**(3) Average delay.** Average delay represents the mean end-to-end time for packets to be delivered from source to destination, including transmission, processing, queuing, and propagation delays. It directly reflects the real-time performance of the SDN network and is given by

$$\text{AverageDelay} = \frac{\sum_{i=1}^{N}(pd_i \times d_i)}{tt} \text{ ms} \tag{13}$$

Where $pd$ is the delay of each transmission path, $d$ is the corresponding successfully transmitted traffic volume, $tt$ is the total transmitted volume, and $N$ is the number of packets within the statistical period.

**(4) Load balance.** Load balance evaluates the uniformity of traffic distribution across network links. The average load $L_{avg}$ and load variance $\sigma^2$ are computed as:

$$L_{avg} = \frac{1}{n}\sum_{i=1}^{n} l_i, \quad \sigma^2 = \frac{1}{n}\sum_{i=1}^{n}(l_i - L_{avg})^2 \tag{14}$$

Where $n$ denotes the number of resources, and $l$ is the load of each resource (e.g., bandwidth utilization). The maximum variance $\sigma^2_{max}$ is then calculated as:

$$\sigma^2_{max} = \frac{(n-1)}{n} \times (L_{max} - L_{avg\_max})^2 \tag{15}$$

Where $L_{max}$ is the maximum capacity of a single link (in Mbps), and $L_{avg\_max} = L_{max} / n$. Thus, the load balance is obtained as:

$$\text{LoadBalance} = 1 - \frac{\sigma^2}{\sigma^2_{max}} \tag{16}$$

During training stage, the Adam optimizer is adopted, together with a StepLR learning-rate scheduler and gradient clipping to ensure training stability. The PPO mixed loss function is employed, comprising the ε-clipped policy loss, mean squared error (MSE) value loss, and an entropy regularization term, thereby balancing policy optimization, value estimation, and exploration. GAE and an experience replay buffer are used for advantage computation and data storage, respectively. Network parameters are iteratively updated in mini-batches, while an adaptive exploration-rate decay strategy dynamically balances exploration and exploitation, enabling efficient learning of SDN routing policies.

All experiments are conducted on a platform equipped with an Intel (R) Xeon (R) Platinum 8251 CPU (3.80 GHz) and an NVIDIA GeForce RTX 4090 GPU, running Ubuntu 20.04.6 LTS, which supports model training, inference, and network simulation. Detailed environment and training parameters are summarized in Table 1.

Table 1 PPO training parameters for SDN routing

| Category | Parameter | Value |
| --- | --- | --- |
| Environment | Network link bandwidth capacity | 1000 Mbps |
| | Baseline traffic generation rate | 100 Mbps |
| | Maximum number of optional routing paths | 8 |
| | Link safety utilization threshold | 0.95 |
| | Controller-switch interaction delay range | (0.001, 0.002) s |
| | Switch flow table maximum capacity (core / aggregation / edge) | 100 / 150 / 200 |
| | Flow entry idle timeout (core / aggregation / edge) | 30 / 20/ 15 s |
| | Default packet size | 1500 Bytes |
| | Traffic statistics time interval | 1.0 s |
| Training | Learning rate for Actor and Critic networks | 5e-05 |
| | Network hidden layer dimension | 16 |
| | PPO policy update clipping coefficient | 0.05 |
| | Entropy reward coefficient | 0.05 |
| | Value loss weight coefficient | 2 |
| | GAE advantage estimation discount factor | 0.96 |
| | Long-term reward discount factor | 0.99 |
| | Training batch size | 32 |
| | Number of update iterations per episode's experience | 15 |
| | Exploration rate decay coefficient | 0.99 |
| | Minimum exploration rate | 0.05 |
| | Gradient clipping threshold | 0.5 |
| | Total number of full training episodes | 3000 |

The overall training procedure of the photonic spiking PPO algorithm is summarized in the pseudocode presented below.

```
Algorithm 1 Spiking-PPO Training Procedure for SDN Routing
Require: Environment 𝔼, max episodes T, batch size N, discount factor γ, GAE
coefficient λ, PPO clip threshold ε_{clip}, entropy coefficient λ_{ent}, value loss coefficient
λ_{val}, update epochs K, exploration decay rate λ_{exp}, min exploration rate ε_{min}
Ensure: Optimized spiking Actor network π^{snn}, Critic network V
 1: Procedure Spiking-PPO training
 2:   Initialize π^{snn}, V, Adam optimizers replay buffer 𝔹,
 3:            exploration rate ε = 1.0, reward history 𝒫 (window = 20)
 4:   for t ← 0 to T−1 do
 5:     Reset 𝔼, get initial state s, episode_reward ← 0
 6:     while 𝔼 not terminated do
 7:       d_a ← 𝔼.get_action_dim(); if d_a = 0 or s = ∅ then break
 8:       a, log_p, v, mask ← π^{snn}.select_action(s, d_a, ε)
 9:       s', r, done, info ← 𝔼.step(a)
10:       Store (s, a, log_p, v, r, done, mask) in 𝔹;
11:       episode_reward += info.raw_reward; s ← s'
12:     end while
13:     Extract batch from 𝔹; v_{term} ← V(s_N)
14:     𝒜, 𝒫 ← GAE (r_1^N, v_1^N, done_1^N, v_{term}, γ, λ);
15:     𝒜 ← (𝒜-𝒜.mean())/(𝒜.std()+1e-8)
16:     for _ ← 0 to K−1 do
17:       Shuffle batch indices (fixed seed)
18:       for start ←0 to N step N do
19:         s_b, a_b, log_{pb}, 𝒜_b, 𝒫_b, mask_b ← batch [start: start + N]
20:         logits ← π^{snn} (s_b).masked_fill(~mask_b, -1e9)
21:         v_{pred} ← V(s_b); dist ← Categorical(logits)
22:         log_{p_new} ← dist.log_prob(a_b)
23:         loss_{policy} ← -min(exp(log_{p_new} - log_{p_b}) × 𝒜_b;
24:         clamp(..., 1 - ε_{clip}, 1 + ε_{clip}) × 𝒜_b).mean()
25:         loss_{total} ← loss_{policy} + λ_{val} × MSE(v_{pred}, 𝒫_b)
26:                - λ_{ent} × dist.entropy().mean()
27:         Update π^{snn} / V via backprop (gradient clipping)
28:       end for
29:     end for
30:     Adjust λ_{exp} by 𝒫's reward growth; ε ← max(ε_{min}, ε × λ_{exp})
31:   end for
32: end procedure
```

## 3. Performance evaluation and experimental results

### 3.1 Training convergence and metrics

Figure 3(a) illustrates the training reward evolution of the photonic spiking PPO algorithm, where "Raw" denotes the original reward and "MA (100)" represents the 100-epoch moving average. The reward increases rapidly during the early training stage and stabilizes after approximately 1000 epochs, indicating effective policy learning and convergence toward high-quality routing decisions. Figure 3(b) shows the corresponding training loss, which exhibits large fluctuations initially but quickly converges to a low level, demonstrating stable and well-behaved optimization. Figures 3(c)–(f) compare the network performance of photonic spiking PPO with the classical Dijkstra algorithm under increasing traffic load. As the base rate increases from 40 Mbps to 200 Mbps, photonic spiking PPO consistently achieves higher throughput, reaching nearly 6000 Mbps, while Dijkstra shows a more limited growth. Meanwhile, photonic spiking PPO maintains a higher load balance across all traffic levels, indicating more uniform traffic distribution and improved congestion avoidance. In terms of reliability and latency, photonic spiking PPO exhibits a significantly lower packet loss rate than Dijkstra, even under high-load conditions, and maintains a relatively stable average delay below 20 ms. In contrast, the delay of the Dijkstra algorithm increases sharply with traffic load, exceeding 80 ms at 200 Mbps. These results demonstrate that photonic spiking PPO can effectively optimize routing paths to improve throughput, balance network load, reduce packet loss, and ensure low-latency transmission.

Overall, the results in Fig. 3 confirm that the proposed photonic spiking PPO algorithm achieves stable convergence during training and consistently outperforms the Dijkstra algorithm across multiple key performance metrics, validating its effectiveness for intelligent routing optimization.

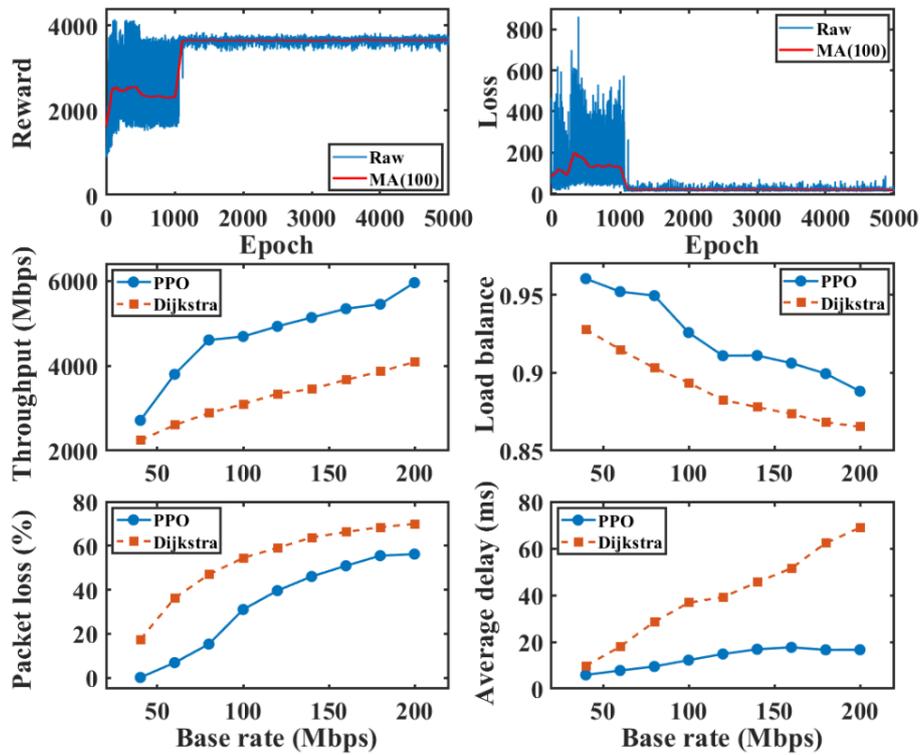

Fig. 3 Training reward and loss curves of the photonic spiking PPO algorithm. Performance comparison between the PPO algorithm and the Dijkstra algorithm, including throughput, load balance, packet loss rate, and average delay.

**3.2 Hardware-software collaborative computing performance**

The SPGD optimization process is shown in Fig. 4(a). To ensure compatibility with PSNN chips, the exported weights are clipped, with negative weights and bias terms removed. The corresponding hardware training results for mapping the hidden-layer linear weights onto the MZI mesh chip are shown in Fig. 4(b). The fidelity of weight mapping is evaluated by computing the similarity between the trained digital weights A and the hardware-mapped weights B,

using the similarity metric defined as follows:

$$a = flatten(A), b = flatten(B), cf = \frac{a \cdot b}{\|a\|\|b\|} \tag{17}$$

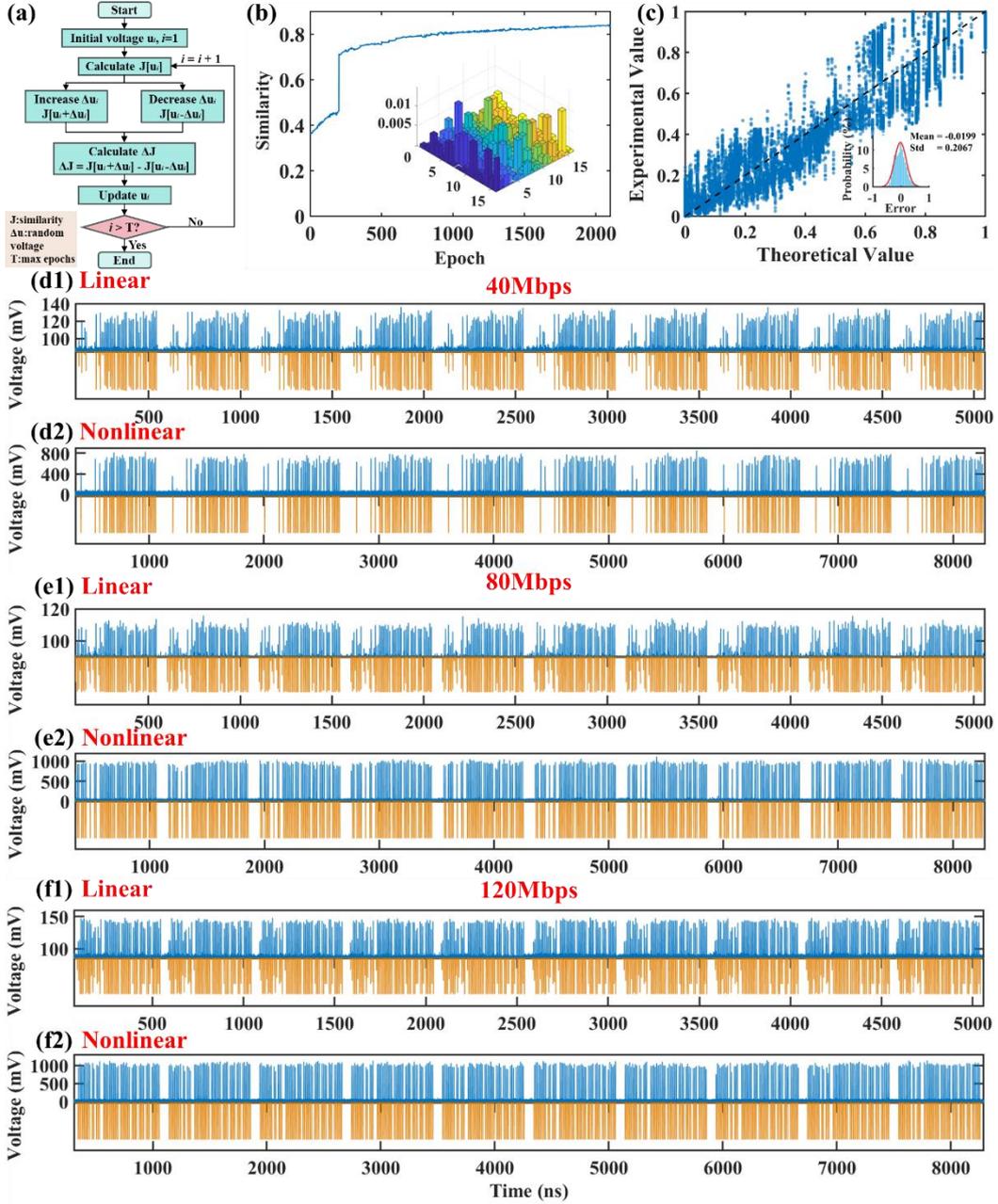

Fig. 4 (a) Optimization process of the SPGD algorithm; (b) MZI training curve; (c) Normalized theoretical and experimental results; Experimentally measured (d1) linear and (d2) nonlinear outputs at a basic rate of 40 Mbps; Experimentally measured (e1) linear and (e2) nonlinear outputs at a basic rate of 80 Mbps; Experimentally measured (f1) linear and (f2) nonlinear outputs at a basic rate of 120 Mbps.

The results show that the similarity approaches 0.80 after approximately 800 training epochs and finally reaches 0.84 after 2100 epochs. Figure 4 (c) presents the normalized theoretical and experimental results, as well as the error distribution of 16 channels (time step T=1) when using 1920 state-action pairs and a total transmitted data volume of

30720. The error probability distribution curve indicates that the errors are mainly concentrated around 0 and exhibit characteristics consistent with the normal distribution. The experimental results are highly consistent with the theoretical predictions, demonstrating that the system maintains excellent stability and accuracy during multi-channel data transmission. Figure 5 shows the retraining results during the hardware-aware software fine-tuning stage. The fine-tuning convergence speed is comparable to the baseline software training (raw reward rises rapidly and stabilizes after about 1000 epochs; loss drops quickly and stays low), proving the stability of the photonic spiking PPO. Though the reward slightly decreases, key metrics (throughput, load balance, packet loss rate, average delay) of the retrained photonic spiking PPO still outperform the classic Dijkstra algorithm, confirming its effectiveness in routing optimization.

Figures 4(d1)-(d2), 4(e1)-(e2), and 4(f1)-(f2) correspond to the experimental results for Channel 10, Channel 3, and Channel 8 under basic traffic rates of 40, 80, and 120 Mbps, respectively. For all three traffic rates, the MZI-based linear computation results exhibit close agreement with the algorithmically calculated target values, indicating high-fidelity optical matrix-vector multiplication across different channels and load conditions. For the 40 Mbps case (Channel 10), as shown in Fig. 4(d2), four overactivation events are observed in the nonlinear activation outputs of the DFB-SA array at 2039.7 ns, 2851.89 ns, 4463.99 ns, and 5264.08 ns. The corresponding layer-wise error rate is 0.5371%, while the final inference accuracy reaches 99.38%; For the 80 Mbps case (Channel 3), only a single overactivation event occurs at 4376.49 ns, as shown in Fig. 4(e2), resulting in a reduced layer-wise error rate of 0.4785% and a final inference accuracy of 100%; For the 120 Mbps case (Channel 8), the nonlinear activation outputs of the DFB-SA array in Fig. 4(f2) are highly consistent with the target activations, with the lowest layer-wise error rate of 0.3516% and a final inference accuracy of 100%. We found that at these overactivation moments, the signal power injected into the DFB-SA laser is close to the activation threshold and is more sensitive to unavoidable disturbances and noise in the system. The reason for the overactivation of the DFB-SA is that the actual input optical power is slightly higher than the actual threshold at the target activation moments.

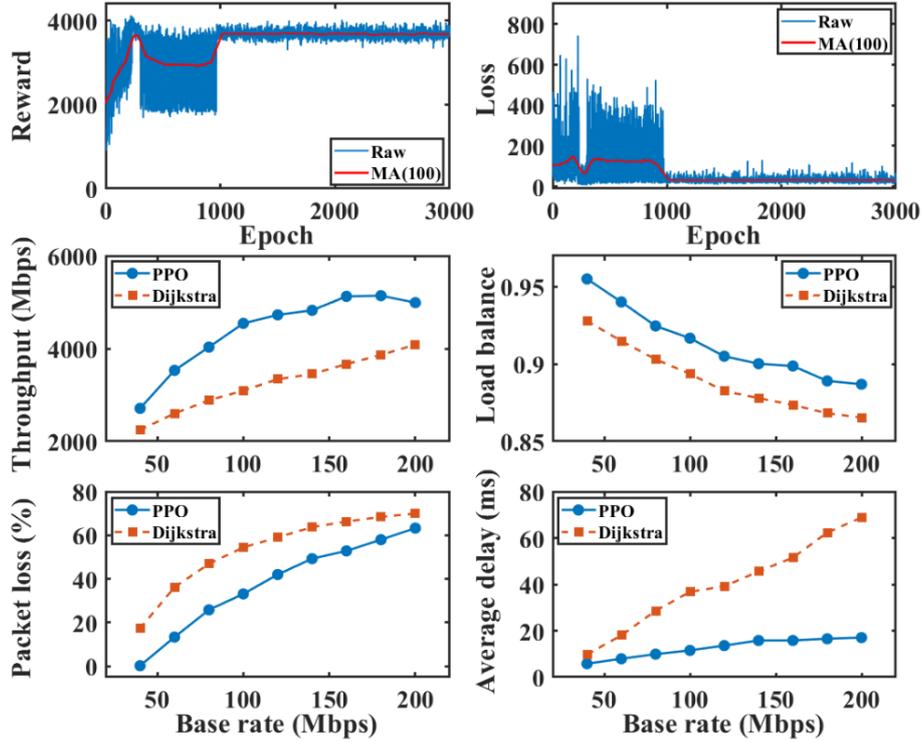

Fig. 5. Training reward and loss curves, as well as throughput, packet loss rate, average delay, and load balance of the photonic spiking PPO network during hardware-aware software fine-tuning.

## 4. Discussion

This section evaluates the scalability of the proposed photonic spiking reinforcement learning–based routing method from two aspects: (1) increasing the SNN hidden-layer size (MZI-mesh scale) and (2) enlarging the network topology (fat-tree with larger k). Training dynamics and routing performance are reported, with comparisons against the classical Dijkstra algorithm.

**Hidden-layer size (128×128 MZI mesh).** Motivated by prior demonstrations of MZI meshes up to 128 ports [28], we evaluate a 128×128 hidden-layer configuration. As shown in Fig. 6, the raw reward increases rapidly during early training, converging to approximately 4000 within a small number of iterations and remaining stable thereafter, indicating efficient learning at this scale. The loss exhibits only brief early fluctuations before quickly converging to a low level, reflecting stable optimization. Across all routing metrics, photonic spiking PPO consistently outperforms Dijkstra. Compared with the 16×16 setting in Fig. 3, the 128×128 model converges faster, while the smaller model shows slightly better late-stage stability due to more extensive early exploration. Nevertheless, both configurations achieve comparable routing performance after convergence, suggesting that the hidden-layer size can be flexibly selected by trading off hardware cost, convergence speed, and training stability.

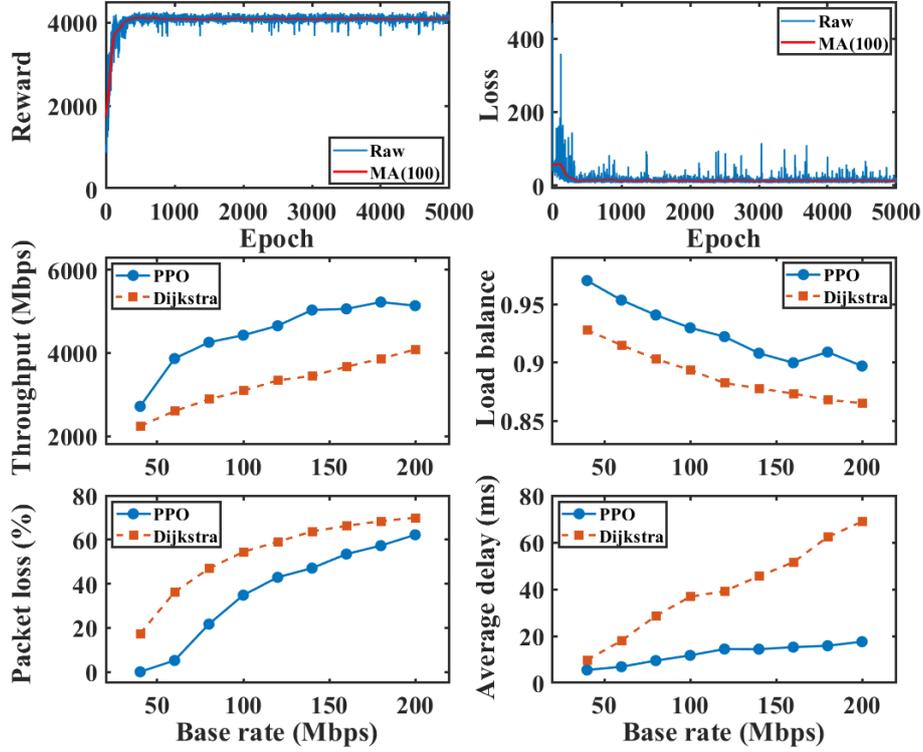

fig. 6 Training reward and loss curves of the photonic spiking PPO algorithm when the hidden layer scale is 128×128, and performance comparison between the PPO algorithm and the Dijkstra algorithm.

**Fat-tree network scale (k = 6).** To assess scalability at the network level, we further evaluate a fat-tree topology with k = 6, comprising 54 hosts and 45 switches. The number of traffic pairs is set to 128, each with a base rate of 50 Mbps during training, while the evaluation phase follows the same traffic-rate configuration as the k = 4 case. As shown in Fig. 7, the reward converges after approximately 600 iterations and stabilizes near 9000, indicating that photonic spiking PPO can rapidly learn a stable routing policy in larger topologies. In terms of performance, photonic spiking PPO achieves a peak throughput of around 12,000 Mbps—approximately doubling that of Dijkstra—benefiting from its adaptive link scheduling capability. For load balance, both methods perform comparably, while photonic spiking PPO consistently achieves lower packet loss rates and average delays. These results demonstrate that the proposed method maintains clear performance advantages and strong generalization as the network scale increases.

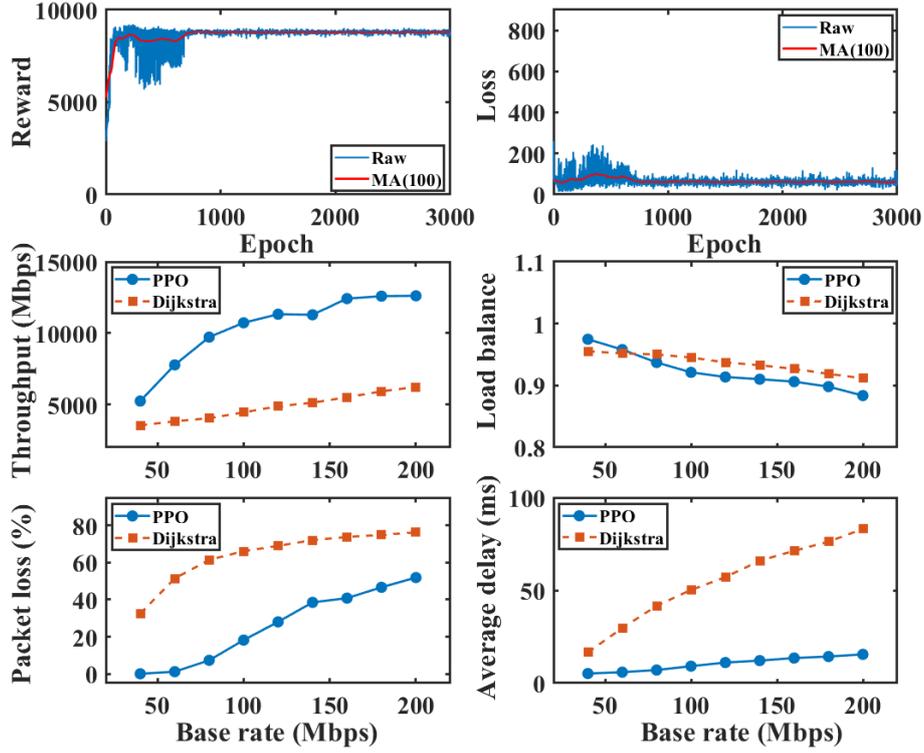

Fig.7 Training reward and loss curves of the photonic spiking PPO algorithm when the fat-tree topology k=6 and the number of traffic pairs is 128, and performance comparison between the PPO algorithm and the Dijkstra algorithm.

**Scalability.** The photonic spiking reinforcement learning chip proposed in this scheme can be scaled to larger size. Prior work shows that, the silicon photonic MZI chip can be expanded to 128 × 128 [28,29], and DFB-SA laser arrays can be scaled to 150 channels [30]. These advances suggest that the integrated PSNN chip can be further enlarged to support intelligent routing in more complex scenarios.

## 5. Conclusions

We propose and implement an SDN routing optimization method based on a photonic spiking PPO algorithm and develop a tailored photonic spiking PPO framework for SDN routing, including well-defined state, action spaces and a multi-objective reward function. This design improves adaptability in dynamic network environments where traditional routing methods are often suboptimal. The approach is validated on a representative fat-tree topology. Experimental results show that, compared with the classical Dijkstra algorithm, photonic spiking PPO achieves clear gains in throughput, packet loss rate, average delay, and load balance, with more pronounced improvements under high network load, demonstrating practical value in complex scenarios. Moreover, we integrate the spiking Actor network with an MZI-based photonic linear computing chip and a DFB-SA–based photonic spiking neuron chip, enabling hardware-software collaborative computing for low-latency, energy-efficient photonic inference. This work bridges photonic neuromorphic computing and communication networks, broadens the application scope of PSNNs, and offers a promising ultra-low latency path for future routing optimization in space–air–ground integrated networks, large-scale data centers, and large-scale satellite networks.

## Acknowledgements

We are grateful for financial supports from the National Natural Science Foundation of China (No.62535015, 62575231), the Fundamental Research Funds for the Central Universities (QTZX23041) and Xidian University Specially Funded Project for Interdisciplinary Exploration (TZJH2024009).

## Author contributions

S Y Xiang: conceptualization, simulations, funding acquisition, revision; Y H Chen: experiments, visualization, measurements, data curation, writing-original draft preparation; L Zheng: methodology, validation, data curation; Z C Tu, X T Zeng, M T Yu, S Wang: experiments, measurements, visualization; Y H Zhang, X X Guo: investigation, data curation, revision; W T Pan, Y Hao: supervision.

## Competing interests

The authors declare no competing financial interests.